# TEX (TEst stand for X-band) at LNF

C. Di Giulio\*, F. Cardelli, S. Pioli,  D. Alesini, M. Bellaveglia, S. Bini, B. Buonomo, S. Cantarella, G. Catuscelli, M.Ceccarelli, R. Ceccarelli, M. Cianfrini, R. Clementi, E. Di Pasquale, G. Di Raddo, R. Di Raddo, A. Falone, A. Gallo, G. Latini, A. Liedl, V. Lollo, G. Piermarini, L. Piersanti, S. Quaglia, L. A. Rossi, L. Sabbatini, G. Scarselletta, , M. Scampati, S. Tocci, R. Zarlenga,

INFN-LNF, [00044] Frascati, Italy

*Abstract—* **TEX facility if commissioned for high power testing to characterize accelerating structures and validate them for the operation on future particle accelerators for medical, industrial and research applications. At this aim, TEX is directly involved in the LNF leading project EuPRAXIA@SPARC_Lab.**

**The brief description of the facility and its status and prospective will be provided.**

## I. Introduction

The project EuPRAXIA@SPARC_LAB aims at constructing a FEL radiation source based on RF LINearACcelerator combined with a plasma module [1,2].

The project is in a preparatory phase of the TDR for this machine and it's one of the pillars of the European Project EUPRAXIA that has been included in the ESFRI 2021 Roadmap. At present this is the only on-going XFEL project worldwide that is based on an entirely new concept aiming at miniaturization of the machine. The new building that will host the new Facility at LNF is under executive design phase and it has obtained authorization for building construction.

### A. The RF LINAC design

The RF baseline of the LINAC is the X-band technology, and it will be used to produce high brightness electron beams up to 1GeV.

The beam will be either injected in the FEL undulators or used to drive a plasma cell to boost the energy further before the undulators. The S-band (2.856 GHz) injector composed by a photocathode RF Gun and 1x 3m TW S-band structure and 3x 2m Travelling Wave (TW) S-band structures.

The X-band (11.994 GHz) booster will be composed by 16x TW structures, 0.9 m accelerating structures with a nominal gradient of 60MV/m.

## II. The TEst stand for X-band

To investigate and test the reliability of the X-band technology for the realization of this booster, all the components, from the source to the accelerating sections, must be tested at the nominal power and working conditions.

As first step in Dec. 2017 an agreement with CERN is signed by INFN to collaborate in the X-band activities. Therefore, the first step was to install a copy of the CERN X-band station at LNF. It was in LNF building #7, very close to the SPARC_LAB area, formerly used for testing and conditioning of the DAFNE RF power plants and cavities.

The X-band test facility, called TEX [3], has been final commissioned last year (2022) in our laboratory and is currently in operation [4]. Different activity was necessary to do for the TEX realization, the first was the decommissioning of the old Decommissioning of the old RF Bunker used to test the DAFNE Cavity. After that a phase of design and commissioning of the new bunker, control room, computing room, storage and auxiliary system (logistics, fluids, safety, radioprotection system…) was necessary to be compliant with the request to the all the Authorities to have their authorizations.

During this phase the modulator k400 procurement and installation of the klystron and components provided by CERN was done and all the other RF components was installed for the SAT.

## III. TEX Facility components:

The TEst-stand for X-band (TEX) is co-funded by Lazio region in the framework of the LATINO project (Laboratory in Advanced Technologies for INnOvation). The setup has been done in collaboration with CERN and it will be also used to test CLIC structures.

It provide all the challenges to build an accelerator with all the subsystems that are summarized in the following sections.

### A. The RF Source

The RF power source is a CPI VKX8311 klystron can generate 50 MW RF pulses, with 1.5 $\mu m$ pulse length at 50 Hz. The klystron parameters and a picture of the klystron during the installation was shown in Fig. 1.

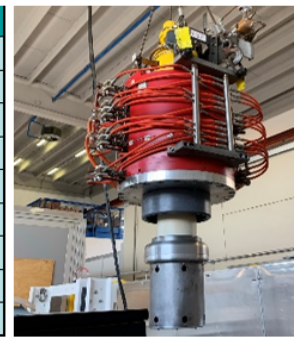

| Typical Operating Parameters | | |
|---|---|---|
| Item | Value | Units |
| Beam Voltage | 410 | kV |
| Beam Current | 310 | A |
| Frequency | 11.994 | GHz |
| Peak Power | 50 | MW |
| Ave. Power | 5 | kW |
| Sat. Gain | 48 | dB |
| Efficiency | 40 | % |
| Duty | 0.009 | % |

Figure 1 On the left the klystron parameters and its installation in the modulator on the right.

The klystron is powered by a modulator model k400 provided by the ScandiNova company. The required stability of the klystron voltage was verified during the SAT to be at level of 10 ppm.

The klystron curve was verified respect the curve provided by CERN and CPI.

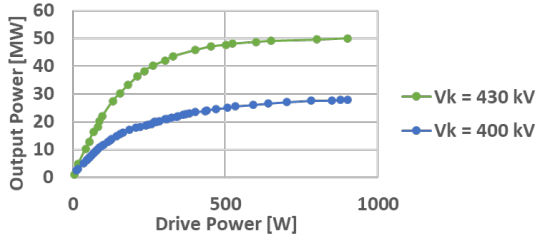

Figure 2 The klystron curve measured a TEX at two different klystron voltage.

The klystron curve was verified respect the curve provided by CERN and CPI for two different klystron voltage at 430 kV and at 400 kV.

B. *The LL RF Source*

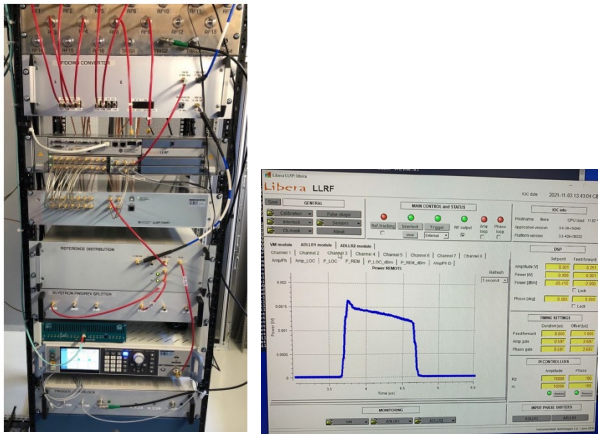

Figure 3 On the right the LLRF rack from the RF source in the lower part to the Libera LLRF and to the upconverter in the higher part where and the RF cable starting point to the klystron and bunker elements. On the left the RF signal mesured by the system.

A The Low-Level RF System The X-band LLRF systems was developed starting by the available on the market S-band system, the Libera LLRF, manufactured by Instrumentation Technologies, whose features and performance have been already reviewed for ELI-NP and SPARC_LAB for a similar architecture [5].

The system shown in figure 3, has been adapted to work at 11.994 GHz by LNF RF team developing different components:

- a reference generation and distribution system able to produce coherent 2.856 GHz S-band and 11.994 GHz X-band references.
- an X-band up/down converter.

Promising results have been obtained, with a measured amplitude and phase jitter of the klystron forward output power of 0.04 % and 20.7 fs respectively. This system does not seem the optimal choice for an X-band based LINAC for various discussed in the reference [5].

C. *The control system*

TEX has selected a standardized, field-proven controls framework, EPICS, which was originally developed jointly by Argonne and Los Alamos National Laboratories. Currently we proceed in the integration of all the hardware equipment present at TEX. IOCs and support modules for any family of device shown in Table 1, has developed, or acquired from repositories, as. An instance of EPICS Archiver Appliance from SLAC to handle the data storage of the facility.

| Device Family |
| --- |
| ScandiNova [6] RF Modulator |
| LiberaLLRF |
| Microwave Amps. [7] RF Driver |
| Pfeiffer Vacuum Gauges |
| Agilent Vacuum Ion-Pumps |
| Timing System |
| RTD sensors |
| SMC Chiller |
| Fluid Plant PLC |
| Faraday Cups |
| Backhoff Motors |
| Basler Camera |
| Magnets |
| Machine Protection System |
| Personnel Safety System |

Table 1 Devices integrated in the TEX control system.

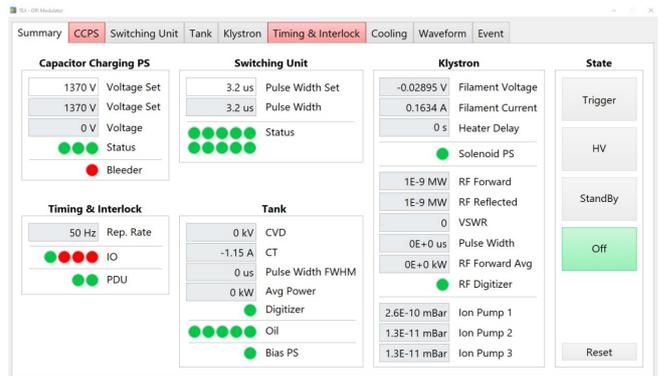

Figure 4 Modulator User Interface.

About User Interfaces (UI) we involved different solution reliable state-of-art graphic interfaces tool to simplify maintenance and improve user experience, where the control panel for modulator operators [6] is shown in Fig. 4.

D. *The building refurbishment*

From the original bunker for the test of the RF DAΦNE cavity a deep refurbishment was necessary to respect the actual safety law for the building.

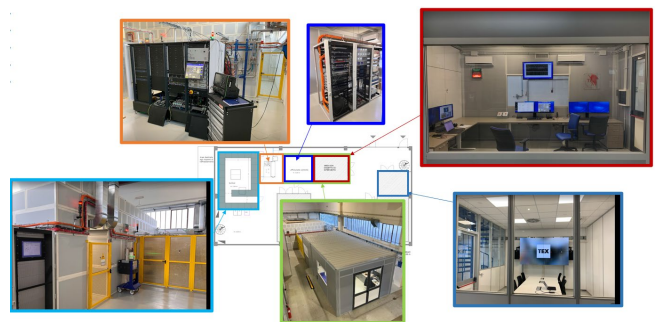

Figure 5 The building refurbishment, from the left the new bunker the modulator area, the computing center the control room and the new meeting area are shown.

In fig. 5 the concrete bunker, the modulator are and the control room are shown. The computing center and a new meeting room are also shown over imposed at the building schema.

*E. The safety system*

At the Safety life-cycle assessment based on statistical methods for risk reduction. The system is complianct with: IEC-61508 standard on "Functional Safety"  the NCRP reports 88 on "Radiation Alarms and Access Control Systems" and the ANSI reports 43.1 on "Radiation Safety for the Design and Operation of Particle Accelerator" .

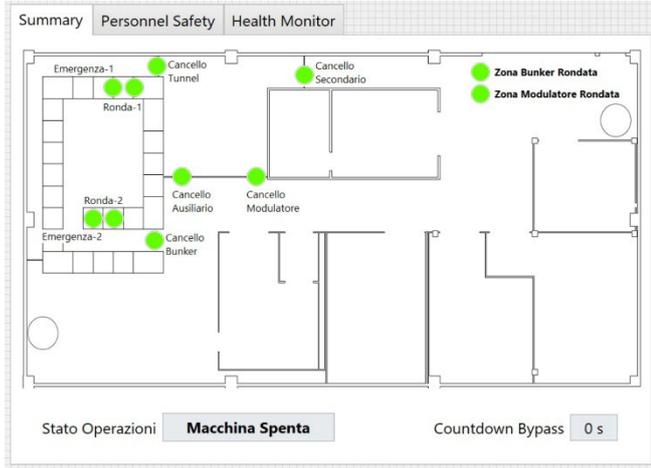

Figure 6 The TEX safety system user interface.

For this reason, we developed Personnel Safety System (PSS) and Machine Protection System (MPS) through FPGA based devices [7, 8] to provide hard-wired, fast and reliable protection to TEX. The safety control rack is shown in figure 5 near the bunker entrance made by the yellow gate and in figure 6 the interlock on the entrances are shown and the other tab for the safety system control user interface are available to the operators.

*F. The vacuum system*

At the The Vacuum system in TEX was provided by Accelerator Division Vacuum Service with the support of Mechanical Service. Since the Device Under Test (DUT) could change rapidly in a test facility, a lot of efforts is required to define vacuum operation procedure for the X band components (cleaning, backing etc..).

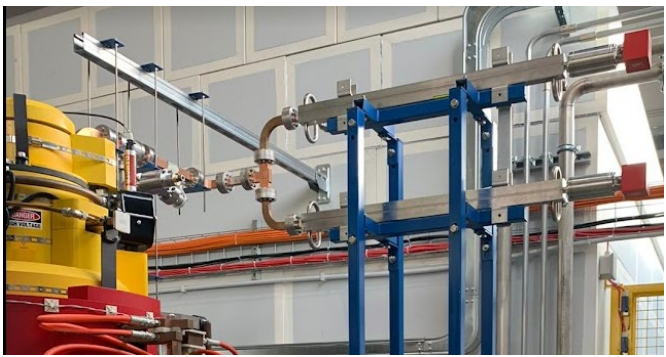

Figure 7 Stainless steel RF load test setup at TEX.

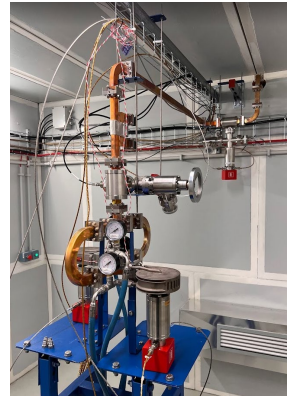

Figure 8 The 3D printed Titanium spiral load setup for the RF high power test at TEX.

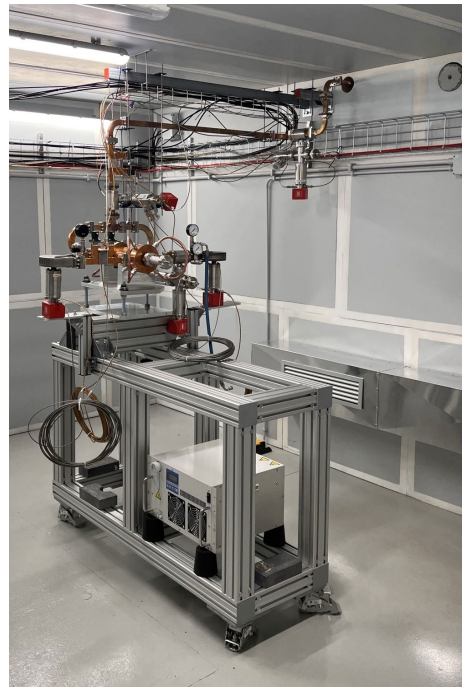

Figure 9 The T24 CLIC accelerating structure, connected with Titanium spiral loads setup ready to be  tested in RF high power at at TEX.

In the last 6 months we change different setups from the configuration shown in figure 7 to the configuration of figure 8 to finish with  RF power  test configuration shown in figure 9.

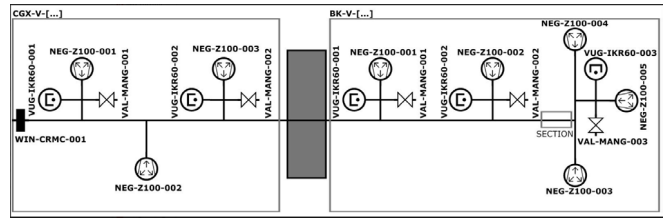

Figure 10 The TEX Vacuum schema shown in Vacuum Control Panel to the operators in control room.

The control of the vacuum quality is provided by the TEX Control System and the Vacuum Status Control panel is available to the operators as the vacuum gauge value and vacuum level value deducted by the ion pump current and is shown in figure 10.

## G. The fluids system

TEX cooling system is based on the EuPRAXIA@SPARC_LAB project guidelines developed by Technical Division at LNF. To reduce the facility opex, the system doesn't use chiller or cooling tower but it's based on adiabatic dry cooler and two independent circuits.

The secondary cooling system keeps the modulator temperature setpoint (28°C) mixing cold water from the dry cooler and the modulator exhaust water.

The primary system cools the water through the adiabatic dry cooler with minimization of electrical cost and water consumption.

A demineralization system maintains the water conductivity at the level required by the system components (< 1μS/cm).

The cooling circuit test for the RF components is performed with a Peltier chiller with thermal stability designed better than 0,03°C.

A PLC based system controls all the cooling system variables and adjusts parameters actuator to maintain the process variable at the setpoint value. Remote control and operation is allowed by a SCADA system.

The bunker thermalization and ventilation is provided by a AHU installed on the bunker rooftop and will be controlled by the same SCADA as the cooling .

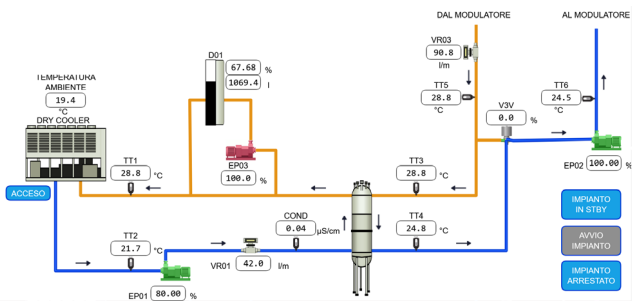

Figure 11 TEX cooling system PLC control interface

## IV. TEX ACTIVITY:

All of the components, from the source to the accelerating sections, must be tested at the nominal power and operating conditions in order to examine and evaluate the reliability of the X-band technology. In the next paragraphs we describe the activities started from the TEX facility commissioning and the prospective of the TEX facility for the next years.

### A. The RF conditioning of waveguides and Spiral Loads RF

A conditioning run of the waveguide distribution line terminated on two 3D printed Titanium Spiral Load has been completed in February 2022.

A final peak power of 42 MW with 250 ns pulse length at 50 Hz has been reached in 3 weeks. The test has been interrupted to perform some civil work in the building.

The FWD klystron power has been gradually increased with an automatic conditioning routine integrated in the control and machine protection system [9].

### B. The TEX next activities

The first test will be to test the prototype of the accelerating structure following the design shonw in figure 12.

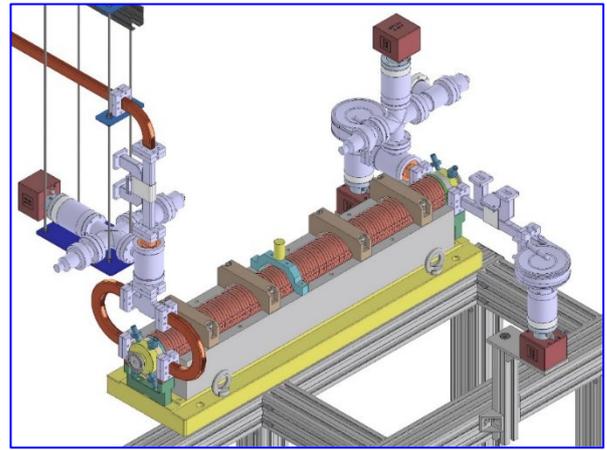

Figure 12 The EuPRAXIA@SPARC_LAB accelerating section design layout at TEX.

A BOC pulse compressor will be integrated in the layout to increase the available power and will be provided by PSI.

Study of design on Pump. Unit, Dir Coupler, Loads, Mode Converter, BOC and section is ongoing. All the elements of the baseline design of the X band module of the EuPRAXIA@SPARC_LAB will be procured by in house or in outsourcing and it will be tested at TEX in the next years [10-12].

### C. The TEX prospective

Guided by the aims to identify the best solution for the EuPRAXIA@SPARC_LAB project, an investment was done by the INFN to test new RF sources available on the market:

- a new modulator with a klystron model CANON E37119 (nominal value 25MW 400 Hz) is scheduled to be installed in 2024;
- a new High Efficiency klystron provided by CPI VKX8311HE it's ordered, and it will be scheduled to be delivered in 2025 at TEX ad it will be installed in the actual k400 TEX modulator.

| Parameter | Unit | Canon E37119 | CPI VKX8311HE |
|---|---|---|---|
| Frequency | MHz | 11994 | |
| Vk beam voltage | kV | 318 | 415 |
| Ik cathode current | A | 197 | 201 |
| Peak drive power | W | 500 | |
| Peak RF output Power | MW | 25 | 50 |
| Average RF output power | kW | 15 | 7,5 |
| Modulator Average power | kW | 75,2 | 25 |
| RF pulse length | us | 1,5 | |
| Repetition Rate | Hz | 400 | 100 |
| Gain | dB | 47 | 50 |
| Efficiency | % | 40 | 55 |

Table 2 Future X band klystron design parameter for the RF source study

The design characteristics of the klystrons is shown in table 2 and a design of the future layout of the TEX facility with both X band RF source is shown in figure 13.

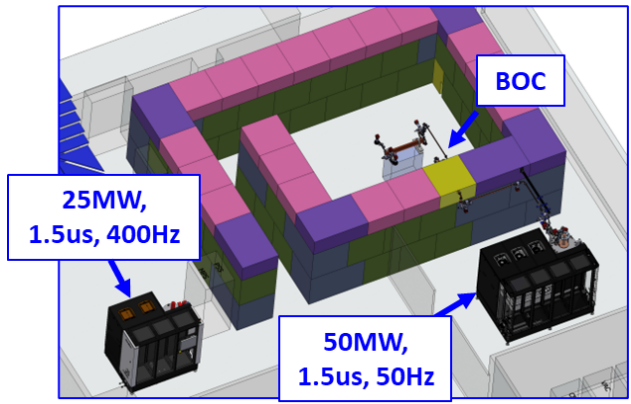

Figure 13 TEX facility design of the new RF power soruces

## V. CONCLUSIONS

The EuPRAXIA@SPARC_LAB is the next INFN-LNF project. The TEX (Frascati Test stand for X-band): is the facility to test all RF components and X band prototypes at the nominal power/gradient. It has been commissioned and in operation. The test of the T24 CLIC structure test is ongoing.

A design and procurement of X-band RF components of for EuPRAXIA RF module have been purchased and will be tested at TEX. An upgrade program of the facility RF source and components is ongoing.

## ACKNOWLEDGMENT

We thank, D. Alesini, S. Bini, B. Buonomo, S. Cantarella, R. Clementi, A. Gallo, C. Di Giulio, E. Di Pasquale, G. Di Raddo, A. Liedl, V. Lollo, L. Piersanti, S. Pioli, R. Ricci, A. Vannozzi on behalf of the TEX and EuPRAXIA technical team, INFN-LNF Accelerator Division, Technical Division and all the Laboratory Administrative and Safety Services staff for the support in the facility commissioning.

We thank W. Wuensh, N. Catalan-Lasheras, A. Grudiev, G. McMonagle on behalf of the CLIC and XBOX group for the provides their expertise in X band technologies.